\documentclass[prl,aps,twocolumn,showpacs,nofootinbib,nobibnotes,superscriptaddress]{revtex4}

\usepackage{epsfig}
\usepackage{amssymb}

\begin{document}

\draft

\title{Gamma Ray Burst Constraints on
Ultraviolet Lorentz Invariance Violation}

\date{July 2006, KSUPT-06/4}
\author{Tina Kahniashvili}
\affiliation{Department of Physics, Kansas State University, 116
Cardwell Hall, Manhattan, KS 66506, USA} \affiliation{National Abastumani
Astrophysical Observatory, 2A Kazbegi Ave, GE-0160 Tbilisi, Georgia}
\author{Grigol Gogoberidze}
\affiliation{National Abastumani
Astrophysical Observatory, 2A Kazbegi Ave, GE-0160 Tbilisi, Georgia}
\author{Bharat Ratra}
\affiliation{Department of Physics, Kansas State University,
116 Cardwell Hall, Manhattan, KS 66506, USA}

\begin{abstract}
We present a unified general formalism for  ultraviolet Lorentz invariance violation
 (LV)
testing through electromagnetic wave propagation, based on  both
 dispersion and
rotation measure data. This allows for a direct comparison of the efficacy of different data to constrain LV. As an example we study the signature of LV
 on
 the rotation of the polarization plane of $\gamma$-rays from gamma
ray bursts in
 a LV
model. Here
$\gamma$-ray polarization data can
 provide a strong constraint on
LV, 13 orders of magnitude more restrictive than a potential
constraint from
 the rotation of
the cosmic microwave background
polarization proposed by Gamboa, L\'{o}pez-Sarri\'{o}n, and Polychronakos
(2006) \cite{GLP05}.

\end{abstract}

\pacs{11.30.Cp, 98.70.Rz, 98.70.Vc}

\maketitle
 Lorentz invariance violation (LV)
has been proposed
as a   possible modification
of the standard model of particle physics and cosmology
(for recent reviews see
Refs.~\cite{k05b,m05,jlm05}).
 Various LV mechanisms have been considered,
including those  motivated by phenomenological quantum gravity, string theory,
non-commutative geometry, and through a Chern-Simons coupling
(for a review see Sec.\ 2 of
Ref.\ \cite{m05}). LV  can
influence  particle propagation (the dispersion relation),
result in  rotation of linear polarization (birefringence),
 and affect the interaction of particles
(including resulting in photon decay and vacuum
\v Cerenkov radiation) \cite{jlm05}.
 These effects can be used to probe LV;
for reviews of current and future tests
see Refs.\ \cite{k05b,m05,jlm05}.

The assumed LV mechanism  determines  the kind of measurements
 required to test the model. Here we study frequency-dependent
 Faraday-like rotation of gamma ray burst (GRB) $\gamma$-ray and $X$-ray
photon polarization in the
context of  ultraviolet LV. For discussions of
 such  high energy LV  see Refs.\ \cite{km01,at01,myers}. Refs.\
  \cite{ccgm03,gl05,GLP05}
study a generalized electromagnetism motivated
by this kind of LV.
On the other hand,  LV associated with a Chern-Simons interaction
\cite{rj98,cfj90} affects the complete spectrum of electromagnetic
 radiation, not just the high-frequency part,  and
 induces a frequency-independent polarization-plane rotation
(see Sec.\ 4 of Ref. \cite{shore}).

In this paper we present a general formalism for LV
testing that encompasses  both rotation measure (RM) and photon
 dispersion measure (DM)\footnote{The DM test is based on the LV effect
 of a phenomenological energy-dependent photon speed \cite{a98} or a modified
 electron dispersion relation.  See Refs. \cite{sarkar} for reviews and Refs.
\cite{jlm03,bwhc04,ellis00,MP06} for recent studies of this
effect; related early discussion include Refs. \cite{16'}. (Refs.
 \cite{a98,bwhc04,MP06}
consider LV models in which rotational and translational invariance are
preserved  but  boost invariance is broken.) }
 observations. This formalism is based on   an analogy with
electromagnetic (EM) wave propagation in a magnetized medium, and
extends previous work \cite{cfj90,km01,jlm03,mmu05}.
 We  show that the
Gamboa et al. (GLP), \cite{GLP05},
LV model is more tightly constrained by RM data than by DM data.
The LV model of Myers and Pospelov (MP),  \cite{myers}, can be tightly
constrained by GRB $\gamma$-ray DM and
 RM observations.
The highly-variable $\gamma$-ray flux of energetic GRB photons
 propagating
over cosmological distances make GRBs  a powerful cosmological
probe \cite{a98} (for reviews of cosmological tests involving
GRBs, see Refs.  \cite{piran,m05}). Testing LV through RM
observations of GRB polarization was proposed,
\cite{mitrofanov,jlms04},
 after the reported observation of highly linearly polarized $\gamma$-rays
from
 GRB021206
  \cite{polarimetry};
this measurement has been strongly contested \cite{pol2}. On the other hand, Ref. \cite{Willis} recently presented evidence that the $\gamma$-ray
flux from GRB 930131 and GRB 960924 is consistent with polarization
degree $>35\%$ and $>50\%$
 respectively.
 Since the issue of  polarization of GRB $\gamma$-rays  still remains
 uncertain{\footnote{For a review of models for generating polarized
 $\gamma$-rays from GRBs see Secs.\ V.F and VI.E of Ref.
 \cite{piran}; more recent discussions
include Refs. \cite{dodo}. For discussions of hard $X$-ray and $\gamma$-ray
polarimetry see Refs.  \cite{mb05,costa}.}},
 we also discuss using future $X$-ray RM
observations.{\footnote{
Ref. \cite{X} predicts linearly polarized $X$-rays from flares following
prompt GRB $\gamma$-ray emission. }}
See Refs. \cite{25} for other RM tests.

We first consider the ultraviolet LV model of GLP  \cite{GLP05}.
Breaking Lorentz invariance leads to a modification of the Maxwell equations
\cite{myers,gl05},  and in  vacuum they become
\cite{GLP05}
\begin{eqnarray}
&&
{\bf \nabla } \cdot {\bf B} = 0,~~~~~~~~~~{\bf \nabla} \times {\bf B} = {\bf {\dot E}},
\nonumber
\\
&&{\bf \nabla } \cdot {\bf E} = 0,~~~~~~~~~~{\bf \nabla} \times {\bf E} +
({\bf g} \cdot { \bf \nabla} ) {\bf {\dot E}} = - {\bf {\dot B}}.
\label{rotE}
\end{eqnarray}
Here
 an overdot represents a derivative with respect to conformal
time $t$,  ${\bf g}$ is the LV vector related to the non-zero
commutator of gauge potentials \cite{GLP05},
 ${\bf B}$ is the magnetic field, and
${\bf E}$ is the electric field that couples to matter in the
usual way but is not related to the gauge potential in the usual
way \cite{GLP05}. To account for the expansion of the Universe we
have to specify how  ${\bf g}$ scales in the expanding Universe. In
 conventional  electrodynamics
the expansion of the Universe is accounted for by a
 conformal rescaling of physical quantities,
i.e. ${\bf{B, E}}
\rightarrow {\bf{B, E}}~ a^2$, where $a$ is the scale factor
 \cite{grasso}. Assuming that the GLP model
 is conformally invariant,
the expansion  may be accounted for by
 rescaling ${\bf g}
\rightarrow {\bf g}/a$, while the components of the physical
electric and magnetic field are diluted as $1/a^2$. On the other hand if the GLP model also violates
 conformal invariance, it is due to a small effect and so the
expansion  can be accounted for as above.
 So GLP LV
results in
 only the Bianchi identity being modified.

In this model the equations for EM wave propagation  in
vacuum are
\begin{eqnarray}
\left[ (\omega^2 - k^2) \delta_{ij} -i \omega  k_l \epsilon_{ijl}
 {\bf  k} \cdot
{\bf g}  \right] E_j({\bf k}) = 0,
\label{eq:02}
\\
k_j E_j({\bf k}) = 0. \label{eq:03}
\end{eqnarray}
Here  $\epsilon_{ijl}$ is the totally antisymmetric symbol,
Latin indices denote space coordinates, $i \in (1,2,3)$, $\omega$
is the angular frequency of the EM wave measured today, and
  ${\bf k}$ is the
wavevector. When transforming between position and wavenumber spaces we use
\begin{eqnarray}
   E_j({\mathbf k}) &=&  e^{i \omega t }
\int d^3\!x \,
   e^{i {\mathbf k} \cdot {\mathbf x}} E_j({\mathbf x}, t),
\nonumber
\\
e^{i \omega t} E_j({\mathbf x}, t) &=& \int {d^3\!k \over (2\pi)^3}
   e^{- i {\mathbf k}\cdot {\mathbf x}}
E_j({\mathbf k}).
\nonumber
\end{eqnarray}
The $e^{i{\omega t}}$ prefactor  describes rapidly varying (compared to
the cosmological expansion time)  EM waves.

A linearly polarized wave can be expressed as a superposition of
 left (L)
and right (R) circularly polarized (CP) waves.
 Using the polarization
basis of Sec.~1.1.3 of Ref.~\cite{varshalovich89},
 Eqs. (\ref{eq:02}) become,
for LCP ($E^+$) and RCP ($E^-$) waves,
\begin{equation}
(\omega^2 - k^2 \mp \omega  k^2 {\bf{\hat k}} \cdot
{\bf g})  E^\pm =0.
\label{RL}
\end{equation}
A similar dispersion relation, in a $D$ brane recoil
model, has been obtained in Ref. \cite{ellis00}. To account for the
phenomenological LV of an energy-dependent
 photon speed \cite{shore,m05,myers,jlm05,mitrofanov},
 we add photon-spin-sign-dependent
 $\mp \gamma (k) k^2 E^\pm (k)$ to the left hand
 side of Eq. (\ref{eq:02})
\cite{jlms04}.
 Here
   (Eq. (5) of Ref. \cite{MP06})
\begin{equation}
\gamma (k) = \left( \frac{\hbar k}{\xi
m_{\rm{pl}}} \right)^q,  \label{eq:03a}
\end{equation}
where $m_{\rm{pl}}$ is the Planck mass, $\hbar $
 is Planck's constant,
$\xi $ is a dimensionless constant that determines the LV energy
scale,{\footnote{ In this case the modification of Maxwell
equations does not preserve conformal invariance \cite{mota}.}}
 and $q$ is a model-dependent number.\footnote{Ref. \cite{kaaret} argues that the
 much-studied $q=1$ case is
almost ruled out by Crab nebula $X$-ray polarimetry data.}
 This modification may be viewed as an
effective photon ``mass'' that makes the photon speed
  less (greater) than the low energy speed of light
$c$ for the RCP (LCP) waves.

To keep the formalism  simple we consider
 an EM wave propagating in the ${\bf z}$ direction with
 ${\bf k}= (0,0,k)$,  and with the LV vector  oriented along the ${\bf z}$
axis, i.e., ${\bf g} = (0,0,g)$. Eqs.~(\ref{RL})
lead to the dispersion relations
\begin{equation}
\omega^2 = k^2 [1 \pm \gamma(k) \pm  g \omega ], \label{wk}
\end{equation}
and in this case $E^\pm =(E_x \pm i E_y)/\sqrt{2}$.
We now  draw an analogy with the propagation of a high-frequency
EM wave in a magnetized plasma.{\footnote{This is motivated by the fact
that LV generates an homogeneous magnetic field
\cite{gl05,mota}.
}}
 High-frequency RCP and LCP waves
propagating in the ${\bf z}$ direction
 in an homogeneous  magnetic field directed along the ${\bf z}$
axis obey \cite{KT}
 \begin{eqnarray}
\left(1-\frac{\varepsilon_1}{n^2} \right) E_x(k) -i
\frac{\varepsilon_2}{n^2}E_y(k)&=&0, \label{eq:04}
\\
i \frac{\varepsilon_2}{n^2}E_x(k) +
\left(1-\frac{\varepsilon_1}{n^2} \right) E_y(k) &=&0.
\label{eq:05}
\end{eqnarray}
Here
 $n = k/\omega$ is the refractive index and
$\varepsilon_1$ and
$\varepsilon_2$ are components of  the  electric
 permittivity or dielectric tensor  $\varepsilon_{ij}$,
  \begin{eqnarray}
\varepsilon_1 &=& \varepsilon_{xx}=\varepsilon_{yy}=1+\frac{\omega_{\rm p}^2}{\omega_{{\rm c}}^2 -
\omega^2},\nonumber\\
\varepsilon_2 &=& \varepsilon_{yx}= -  \varepsilon_{xy}=
\frac{\omega_{\rm c}}{\omega}
\frac{\omega_{\rm p}^2}{\omega_{\rm c}^2 - \omega^2},  \label{eq:06}
\end{eqnarray}
where $\omega_{\rm p}$ and $\omega_{\rm {c}}$  are the plasma and
 electron cyclotron angular frequencies (see Sec. 4.9 of Ref.
 \cite{KT}).

In the magnetized plasma case
 an homogeneous magnetic field induces a phase velocity
 difference between LCP and RCP waves and so  causes rotation of the
 polarization plane.
 Also, in this case, the
group velocity of an EM wave differs from $c$ and so
 results in time delay. These two independent DM and RM effects
 can be
expressed in terms of  refractive indices,  $
n_{{\rm L, R }}=(\varepsilon_1  \mp
\varepsilon_2)^{1/2}$, where the sum (lower sign) corresponds to the
RCP wave \cite{KT}.{\footnote{The LCP and RCP EM wave electric fields
  obey
$[n^2 - (\varepsilon_1 \pm \varepsilon_2)]E^\pm=0$ \cite{KT}.
 The basis vectors $({\bf e}^+, {\bf e}^-,
{\bf \hat z})$ satisfy
${\bf e}^\pm \cdot {\bf e}^\mp =1$,
 ${\bf e}^\pm \cdot {\bf e}^\pm =0$, ${\bf e}^\pm ({\bf\hat z}) =
{\bf e}^\mp (- {\bf\hat z})$, and $\pm {\bf e}^\pm = i {\bf \hat z} \times {\bf e}^\pm$ \cite{varshalovich89}. }}
 As a consequence the LCP and RCP wavevectors are
 $k_{\rm{L, R}} = \omega n_{\rm L,R}$.
Both DM and RM effects depend on the
 photon travel  distance $\Delta l$  and are expressed through
 \begin{eqnarray}
\Delta t_{\rm L,R} &=& {\Delta l} \left( 1 - \frac{\partial
k_{\rm L,R}}{\partial \omega} \right),
\label{time-delay}
\\
\Delta \phi &=& \frac{1}{2}(k_{\rm L} - k_{\rm R}) \Delta l.
\label{angle}
\end{eqnarray}
Here $\Delta t_{\rm L,R}$ is the difference between  the
 LCP (RCP) photon travel time and that for a ``photon''  which
 travels at $c$, and
 $\Delta \phi$ is the polarization-plane
rotation angle.

We can rewrite Eqs. (\ref{eq:02}) and (\ref{eq:03})
for the LV case in a form similar to Eqs.
(\ref{eq:04}) and (\ref{eq:05}) for a magnetized plasma.
 Define two dimensionless quantities
 $\varepsilon_1^\pm = 1/(1 \pm \gamma(k))$ and
 $\varepsilon_2^\pm  = - g k^2 /[ \omega (1 \pm \gamma(k))] $.
Experimentally LV is  small so we   simplify by taking
 $\gamma(k) $ and $g\omega$ to be small and work to  linear
order in these quantities. To linear order, $\varepsilon_2^\pm
 \approx - g\omega$ and is independent of the photon spin sign, while
 $\varepsilon_1^\pm \approx 1 \mp \gamma$ and  depends on
the photon spin sign.
The corresponding (L, upper sign) and (R, lower sign)
refractive indices are
$
n_{{\rm L, R}}=(1\pm \gamma \pm g\omega)^{1/2}$.
Both kinds of LV (scalar $\gamma$ and vector $\bf g$) induce  DM
and RM effects.  There are  two different regimes of interest,  when
$\gamma  \gg g \omega $ and
when $\gamma \ll g \omega$.

When  $\gamma(k)  \gg g\omega $,
 as in the MP model \cite{myers},
Eqs. (\ref{time-delay}) and (\ref{angle}) become
\begin{eqnarray}
\Delta t_{\rm{L, R}} &\simeq & \mp \frac{\Delta l}{2} (1+q) \gamma(k),
\label{eq:10} \\
\Delta \phi &\simeq & \frac{\Delta l}{2} \omega \gamma(k).
\label{eq:10a}
\end{eqnarray}
These expressions agree with those obtained earlier in
Refs. \cite{mitrofanov,jlms04,MP06}. DM and RM measurements
can be used to
constrain $\gamma$. DM testing  of LV
through the time delay of GRBs photons  has been widely
discussed (for a recent review
see Ref.~\cite{jlm05}) and so is not discussed
here.

When $\gamma(k)  \ll g \omega$,  as in the GLP model \cite{GLP05},
 Eq. (\ref{angle}) yields (see also the Conclusion
of Ref.~\cite{GLP05}),
\begin{equation}
\Delta \phi \simeq \omega^2 g \frac{\Delta l}{2}, \label{eq:11}
\end{equation}
and  Eq. (\ref{time-delay})
for the time delay gives
\begin{equation}
\Delta t_{\rm{L, R}} \approx \mp g \omega \Delta l.
\label{eq:13}
\end{equation}
DM and RM measurements constrain the value of
$\varepsilon_2$ (or $g\omega$), but the dependence on frequency
is different, with the constraints from the RM test being strongest
for high-frequency waves.

For  ``classical'' Faraday rotation  $\Delta \phi \sim
\omega^{-2}$, \cite{kl96}, and the effect is strongest for
low-frequency waves.
 For  GLP LV
$\Delta \phi \sim \omega^2$
 and the effect is strongest
for high-frequency waves.
  Ref.~\cite{GLP05} suggests using
cosmic microwave background (CMB) polarization data to test GLP LV, as  was previously
proposed to
detect a primordial cosmological magnetic field \cite{grasso,kl96} and
 test for CPT
violation \cite{flxcz06}. We argue below that GRB $\gamma$-rays  polarization
 measurements will give a
much stronger bound
on this kind of  LV.
 On the other hand, lower frequency CMB polarization data
may be used to constrain LV
induced by a Chern-Simons coupling since in this case the RM is
 frequency independent
 \cite{cfj90} (this will complement
the limit obtained from  radio galaxy RM data \cite{cfj90}).

It should be possible to measure a
 $\Delta \phi \sim 10^{-2}$ rad.
For CMB radiation
 with $\omega \sim  10^{11}~{\rm Hz}$ and for
 photon travel distance
$\Delta l \sim 1.3  \times 10^{10}$ y, the RM  GLP LV
 constraint, Eq. (\ref{eq:11}), indicates that one may probe to
 \begin{equation}
g_{\rm {CMB}} \sim   10^{-18}~{\rm Gev^{-1}}. \label{eq:12}
\end{equation}
For  GRB $\gamma$-rays  with
 $\omega \sim
10^{19}~{\rm Hz}$ and $ \Delta l \sim 3 - 5  \times 10^9$ y, even
with less accurate RM data with, say, $\Delta \phi sim 1$ rad, Eq.
(\ref{eq:11}) shows that there is detectable LV down to
\begin{equation}
g_{\rm{GRB}} \sim    10^{-31}~{\rm Gev^{-1}}. \label{eq:12a}
\end{equation}
In the GLP model GRB $\gamma$-ray  data
 can probe 13 orders of magnitude
higher in energy than can CMB data. Note that synchrotron radiation
RM data at $\omega=340$ GHz \cite{marone} from Sagittarius A$^\star$
at $ \Delta l \simeq 2.5 \times  10^4$ y with $\Delta \phi \simeq
0.5$ rad gives the weaker constraint $g_{\rm Sag}\approx 10^{-11}$
Gev$^{-1}$. The polarization data at the optical band from active
galactic nuclei give 8 magnitudes weaker limits than GRB future
data.

To compare the relative efficacy of RM and DM data at probing LV,
we consider the ratios of the same-source DM and RM data LV limits
 for the two characteristic LV quantities
$\xi^{-1}$ and $g$,
\begin{equation}
r_\xi=\frac{\xi^{-1}_{\rm{DM}}}{\xi^{-1}_{\rm{RM}}},
~~~~~~~~
r_g=\frac{g_{{\rm DM}}}{g_{{\rm RM}}}.
\label{rs}
\end{equation}
The constraints on $\xi^{-1}$ in the case when $\gamma \gg
g \omega$ and $k \simeq \omega$
 can be obtained from Eqs. (\ref
{eq:03a}), (\ref{eq:10}), and (\ref{eq:10a}),
\begin{equation}
 \xi_{\rm{DM}}^{\rm{L,R}} = \frac{\hbar\omega}{m_{\rm{pl}}}
\left[\frac{(q+1) \Delta l }{\mp 2 \Delta t_{\rm{L,R}}} \right]^{1/q}, ~~~~
\xi_{\rm{RM}} = \frac{\hbar\omega^{1+1/q}}{m_{\rm{pl}}}
\left[\frac{\Delta l }{2 \Delta \phi} \right]^{1/q}.
\label{xi}
\end{equation}
The constraints on $g$ when $\gamma \ll
g \omega$
 can be obtained from Eqs. (\ref{eq:11}) and (\ref{eq:13}),
\begin{equation}
 g_{\rm{DM}}^{\rm{L,R}} = \mp \frac{\Delta t_{\rm{L,R}}}{\omega \Delta l}, ~~~~
g_{\rm{RM}} = \frac{2\Delta \phi}{\omega^2 \Delta l}.
\label{g-limit}
\end{equation}

We first consider GLP LV where  $\gamma
\ll g\omega$.
Using the GRB $\gamma$-ray parameters
 mentioned above, taking  $|\Delta t_{\rm{L,R}}| =
10^{-4}~{\rm s}$ as the  current
accuracy of time delay data  \cite{MP06}, and
 assuming $\Delta \phi =1$ rad,
$|r_g^{\rm{GRB}}| = \omega |\Delta t_{\rm{L,R}}|/(2 \Delta \phi)
 \sim 10^{14}
$.
So in this case the limit from RM data is
strongest. If one wishes to constrain $g$ using
 GRB DM and CMB RM data, then
$|r_g^\star|=|g_{\rm{DM}}^{\rm{GRB}}|/g_{\rm{RM}}^{\rm{CMB}} \approx
0.2$,
  so both are almost equally good tests for LV.

In the opposite case when $\gamma \gg g\omega$,
if DM and RM data from the same source are used,
\begin{equation}
r_\xi^{\rm{L,R}} =
 \left[\frac{ \mp \omega \Delta t_{\rm{L,R}}}{(q+1) \Delta \phi }
 \right]^{1/q}.
\label{rGRB1}
\end{equation}
Conventionally two cases are considered,  the  linear case
 with $q=1$ \cite{jlm03,MP06}, and
the quadratic case with  $q=2$ \cite{mitrofanov,MP06}. For $q=1$ Eq.
(\ref{rGRB1}) reads for GRB $\gamma$-rays  $ |r_\xi^{\rm{GRB}}| =
\omega |\Delta t_{\rm{L,R}}|/(2\Delta \phi) \sim 10^{14}$. Note that
our $\xi $ is the inverse of the $\xi$ of Ref. \cite{jlm03}
  and coincides with the
 $\xi$ of Ref. \cite{MP06}.
 Using the GRB $\gamma$-ray parameters considered above, we see that CMB
polarization RM data may slightly improve the $\xi$ limit obtained from
GRB $\gamma$-ray DM data \cite{MP06}. The improvement will be much more
significant  if GRB $\gamma$-ray RM data is used
 \cite{mitrofanov,jlm05}. For the $q=1$ MP model \cite{myers}
 $r_g$ and $r_\xi$  are the same order of magnitude;
 i.e., RM data used  for
 frequency $\omega
>  2 \Delta \phi/ |\Delta t_{\rm{L,R}}|$ results in similar limits
on  $g$ and $\xi^{-1}$.
  With $q=1$, as a consequence of
 the frequency dependence $|r_{g, \xi}|\propto \omega$,
 high-frequency data result in more restrictive constraints.
 For the  $q=2$ case,  $r_\xi \propto
 \sqrt{\omega}$ and $r_{g}\propto \omega$,
 so the potential limit on $\xi^{-1}$
 from GRB $\gamma$-ray RM data 
 is 6---7 orders of magnitude better
than that from DM data \cite{mitrofanov}.

In summary, we present a unified general treatment of both
 LV DM and RM tests  by analogy with EM wave propagation in a
 magnetized plasma. This
 treatment
  does not depend on the LV model, and allows
 simultaneous consideration of
 different LV mechanisms. We considered
conventional ultraviolet LV,  i.e. linear MP, quadratic MP, and
GLP models. For these models, RM data provide better limits than DM data,
(the improvement is $\sim 100$ for
 linear MP and GLP LV, $\sim 10$ for quadratic MP LV,
 if $\omega > 100$ kHz),
 and  the improvement increases by using higher frequency
 EM wave data (for an arbitrary MP model $r_\xi \propto
 \omega^{1/q}$ and thus RM test efficacy decreases as $q$ increases).
Future
 $\gamma$- and $X$-ray  RM data from distant objects,
such as GRBs, quasars, or blazars hold great promise for
 testing and strongly constraining LV.

It is our pleasure to thank G. Melikidze for fruitful comments and
suggestions. We also thank G. Gabadadze, A. Kosowsky, I.
Litvinyuk, I. Mitrofanov, A. Sakharov, L. Samushia, S. Sarkar, T.
Vachaspati, and L. Weaver for helpful discussions. T.K. and B. R.
acknowledge support from DOE grants DE-FG02-00ER45824 and
DE-FG03-99EP41093. T. K and G. G. acknowledge partial support from
INTAS grant 06-1000017-9258.

\end{document}